\documentclass[12pt]{article}
\usepackage{geometry,enumerate,amsmath,amssymb}
\usepackage{fullpage}
\usepackage{graphicx}
\newcommand{\be}{\begin{equation}}
\newcommand{\ee}{\end{equation}}

\newcommand{\pauli}{\mbox{\boldmath $\tau$}}
\renewcommand{\hat}{\widehat}

\renewcommand{\epsilon}{\varepsilon}

\font\mybb=msbm10 at 11pt

\def\bb#1{\hbox{\mybb#1}}

\newcommand{\news}{\setcounter{equation}{0}\quad}
\def\ben{\begin{equation}}
\def\een{\end{equation}}
\def\bea{\begin{eqnarray}}
\def\eea{\end{eqnarray}}
\begin{document}
\title{
\begin{flushright}\ \vskip -2cm {\small {\em DCPT-11/13}}\end{flushright}
\vskip 2cm Monopoles in AdS}
\author{
Paul Sutcliffe\\[10pt]
{\em \normalsize Department of Mathematical Sciences,
Durham University, Durham DH1 3LE, U.K.}\\[10pt]
{\normalsize Email: 
 \quad p.m.sutcliffe@durham.ac.uk}
}
\date{July 2011}
\maketitle
\begin{abstract}
Applications to holographic theories have led to some recent interest
in magnetic monopoles in four-dimensional Anti-de Sitter spacetime.
This paper is concerned with a study of these monopoles, using
both analytic and numerical methods. 
An approximation is introduced in which the fields of a 
charge $N$ monopole are explicitly 
given in terms of a degree $N$ rational map. 
Within this approximation, 
it is shown that the minimal energy monopole of charge $N$
has the same symmetry as the minimal energy Skyrmion with baryon number $N$ 
in Minkowski spacetime. Beyond charge two the minimal 
energy monopole has only a discrete symmetry, which is often Platonic.
The rational map approximation provides an upper bound on the monopole 
energy and may be viewed as a smooth non-abelian
refinement of the magnetic bag approximation, to which it reverts under
some additional approximations. The analytic results are supported
by numerical solutions obtained from simulations of the non-abelian 
field theory.
A similar analysis is performed on the monopole wall that
emerges in the large $N$ limit, to reveal a hexagonal lattice
as the minimal energy architecture.
\end{abstract}
\newpage
\section{Introduction}\news
The AdS/CFT correspondence allows the investigation of 
strongly coupled theories by studying classical solutions in the bulk. 
Recently, it has been argued
that interesting phenomena may result, including spontaneous
breaking of translational symmetry, if the bulk Anti-de Sitter (AdS)
spacetime contains
non-abelian magnetic monopoles \cite{BT}. This provides motivation for
a detailed study of $SU(2)$ magnetic monopoles in four-dimensional 
AdS spacetime.

In addition to holographic applications, there are other reasons to
consider monopoles in AdS. In Minkowski spacetime, when the Higgs
field is massless, there is a $4N$-dimensional
moduli space ${\cal M}_N$ of static charge $N$ BPS monopoles. 
In contrast, the attraction in AdS spacetime
should produce a unique minimal energy monopole for each charge $N$
(up to the obvious action of spatial rotations). It is of interest to 
determine the structure and symmetry of the $N$-monopole, which will
map to a unique point in the moduli space ${\cal M}_N$ as the AdS
curvature tends to zero. 

It has been known for some time that there are many similarities
between BPS magnetic monopoles and Skyrmions (for a review see \cite{MSbook}).
However, one important difference is the attractive force between
Skyrmions, producing bound states, in contrast to the
BPS monopole moduli space ${\cal M}_N,$ resulting 
from the absence of static forces between monopoles.
It is therefore expected that studying monopoles in AdS will enhance
the similarities with Skyrmions, since both will share the 
features of attractive forces and bound states.  

Motivated by previous work on Skyrmions \cite{HMS}, 
an approximation is introduced in which the non-abelian fields of 
an $N$-monopole in AdS are written in terms of a degree $N$ rational map
between Riemann spheres. The Yang-Mills-Higgs energy functional
leads to an energy functional on the space of rational maps, which is
precisely the one found in the similar approach to Skyrmions.
This implies that, within this approximation, 
the minimal energy monopole of charge $N$ in AdS
has the same rotational symmetry group as the minimal energy 
Skyrmion with baryon number $N$ and massless pions in
 Minkowski spacetime. 

In particular, this approach predicts that
 the 1-monopole is spherically symmetric and the 2-monopole
is axially symmetric, but for charges greater than two the minimal
energy monopole has only a discrete symmetry group, which is often Platonic.

In Minkowski spacetime the existence of particular BPS monopole 
solutions with Platonic symmetries has been proved \cite{HMM,HS2}
using the Nahm transform \cite{Nahm}, but in the BPS case there
is no energetic preference for these solutions, because of the 
democracy of the moduli space ${\cal M}_N$.   
Although the rational map approximation does not produce any exact solutions,
beyond charge one, it does provide an upper bound on the $N$-monopole 
energy and suggests that in moving from Minkowski spacetime to
AdS, the breaking of the energy degeneracy of the moduli space ${\cal M}_N$
leaves particular symmetric monopoles as the minimal energy solutions.

Monopoles with large charge have been considered previously using
the magnetic bag approximation, both in Minkowski spacetime
\cite{Bo} and in AdS \cite{BT}. The magnetic bag approximation assumes
that inside a spherical bag both the Higgs field and the magnetic field
vanish and outside the bag there is an abelian magnetic field.
The rational map approximation may be viewed as a smooth non-abelian
refinement of the magnetic bag approximation, 
to which it reverts under some additional approximations. 

Simulations of the full non-abelian field theory are performed
to provide numerical solutions for monopoles in
AdS, with charges from 1 to 17. 
These numerical results provide support for both the approximate analytic
treatment using rational maps, and the magnetic bag approximation.
The numerical results also reveal interesting information about
the zeros of the Higgs field. For most of the solutions, there
is a zero of the Higgs field with multiplicity $N$ located at the origin. 
However, for $N=3$ and some further non-minimal solutions with
higher charge, there are $N+2$ zeros of the Higgs field, with
one of the zeros having a negative multiplicity.  
 
By zooming to the Poincar\'e patch in the large $N$ limit, the 
study of magnetic monopoles translates into the investigation of
monopole walls in AdS. A monopole wall \cite{Wa} is a novel domain
wall in which 
the magnetic field along a line perpendicular to the wall 
tends to zero on one side of the wall and to a non-zero constant on 
the other side of the wall.
The fields are periodic in the directions parallel to the wall, with
a non-trivial spatial variation of the energy density and magnetic
field. Monopole walls have infinite energy per unit area in Minkowski 
spacetime, but in AdS this is finite and they can be studied using
a magnetic bag style approximation \cite{BT}.

 The magnetic bag
approximation is too crude to reveal any information concerning the
spatial distribution of the fields parallel to the wall, but a
variant of the rational map approximation, involving
elliptic functions, is refined enough for this
purpose. It suggests that the minimal energy per unit area is
obtained from a monopole wall with a hexagonal lattice.
Numerical simulations are performed that support this conclusion, 
which is again in agreement with the Skyrme model, where a similar
hexagonal lattice exists \cite{BS-lat}.      

Previous studies of monopoles in AdS have been restricted to the 
spherically symmetric single monopole \cite{LS,LMS}, and axially symmetric
monopoles with charges two and three \cite{RT}. 
The computations in \cite{RT} reveal that the energy of the 
charge two monopole is greater than twice the energy of the single
monopole, but this result was incorrectly interpreted to conclude
that monopoles repel and that the axially symmetric 2-monopole is unstable.
The source of the confusion is that 
in Minkowski spacetime the following argument can be applied: if the 
energy of an $N$-soliton is greater than $N$ times the energy of
a single soliton, then the energy of the $N$-soliton can be reduced
by infinitely separating the $N$ constituents. However, in AdS this
argument fails because the energy of a single soliton increases as
it approaches the boundary of AdS. The energy of the single soliton,
which is used in the false comparison, is the single soliton energy 
only when the soliton is located at the origin of AdS. 
At first glance this may appear confusing, given the large 
isometry group, $O(3,2)$ of AdS, but there is an important subtlety
at work here. To consider static solutions, and their associated
energies, requires the selection of a time-like Killing vector. 
This breaks the isometries of AdS and introduces a preferred point
in space, which is denoted the origin. Solitons are attracted towards
the origin, which is consistent with the interpretation of AdS
as gravitational attraction.

\section{Monopoles, rational maps and magnetic bags}\news
In terms of sausage coordinates, the metric of four-dimensional
AdS spacetime may be written as
\be
ds^2=-\bigg(\frac{1+\rho^2}{1-\rho^2}\bigg)^2\,dt^2
+\frac{4L^2}{(1-\rho^2)^2}
\bigg(d\rho^2+\rho^2(d\theta^2+\sin^2\theta \,d\varphi^2)\bigg),
\label{ads}
\ee
where the range of the radial coordinate is $0\le \rho\le 1$ 
and $L$ is the AdS radius, related to
the cosmological constant via $\Lambda=-3/L^2.$
The above choice of coordinates for AdS is motivated by later
numerical investigations, as it is more efficient to have
a finite range for the coordinates.

The Yang-Mills-Higgs action is
\be
S=\int \frac{1}{2}\mbox{Tr}
\bigg(\frac{1}{4}F_{\mu\nu}F^{\mu\nu}+\frac{1}{2}D_\mu\Phi D^\mu\Phi\bigg)
\sqrt{-g}\,d^4x,
\label{action}
\ee
where the gauge potential $A_\mu$ and Higgs field $\Phi$ are both
 $su(2)$-valued. The massless Higgs field transforms in the adjoint 
representation of $SU(2)$ and is subject to the boundary condition
$|\Phi|^2=-\frac{1}{2}\mbox{Tr}\,\Phi^2\rightarrow 1$ as 
$\rho\rightarrow 1.$ 

At a fixed time, 
the Higgs field on the spatial boundary is a map between
two spheres, $\Phi|_{\rho=1}:S^2\mapsto S^2,$ with winding
number $N\in \pi_2(S^2)=\bb{Z}.$  
This winding number is equal to the magnetic charge of the monopole,
in units of $2\pi,$ and may also be written as the integral
\be
N=\frac{1}{4\pi}\int\varepsilon_{ijk}\mbox{Tr}(F_{jk}D_i\Phi)\, d^3x.
\label{charge}
\ee
Static monopoles are critical points of the static energy
associated with the action (\ref{action}), namely
\be
E=-\int \frac{1}{2}\mbox{Tr}
\bigg(\frac{1}{4}F_{ij}F^{ij}+\frac{1}{2}D_i\Phi D^i\Phi\bigg)
\sqrt{-g}\,d^3x.
\label{energy}
\ee
The aim is to find the global minimum of this energy, within each
topological sector given by the positive integer $N.$ 
Note that the above normalizations are
such that in Minkowski spacetime each point of the BPS moduli
space ${\cal M}_N$ is associated with a monopole solution with 
energy $2\pi N.$ 

It is perhaps worth pointing out that although constant
time slices of the metric (\ref{ads}) give the Poincar\'e ball
model of hyperbolic space, the static energy (\ref{energy}),
and hence the static field equation, is not the much studied
system for monopoles in hyperbolic space \cite{At}.
This is because of the extra factor
$\sqrt{-g_{tt}}=(1+\rho^2)/(1-\rho^2)$
hidden in the term $\sqrt{-g}$ from the warp factor in the metric
(\ref{ads}). 

In Minkowski spacetime there is a diffeomorphism, 
preserving the action of spatial rotations,
 between the BPS moduli space ${\cal M}_N$ and a certain
 equivalence class of degree $N$ rational maps between Riemann spheres 
\cite{Ja}. 
The rational map appears as scattering data 
along a radial half-line of an operator constructed from the monopole fields.
As the inverse scattering problem is not tractable
for $N>1,$ there is no explicit formula for the monopole fields
in terms of the rational map.

Motivated by similarities between monopoles and Skyrmions, an 
explicit but approximate description of Skyrmions was introduced 
in terms of rational maps \cite{HMS}. Continuing the theme 
of an interplay between monopoles and Skyrmions, the rational map
approximation for Skyrmions will now be adapted back to the 
monopole context, to provide approximate fields for monopoles in
AdS.    

A rational map between Riemann spheres, $R(z),$ is simply a ratio 
of two polynomials in a complex variable $z.$ 
The two polynomials are required to have no common roots, and 
the degree of the rational map is the largest of the degrees of 
the two polynomials. To make the connection to monopoles (or Skyrmions)
the Riemann sphere coordinate $z$ is related to the angular 
space coordinates $\theta$ and $\varphi$ via standard stereographic
projection, that is
$z=e^{i\varphi}\tan(\theta/2).$ 

The Riemann sphere coordinate, $R,$ on the target space of the rational map,
is related to a three-component unit vector ${\bf n}$ by inverse stereographic
projection,
\be
{\bf n}=\frac{1}{1+|R|^2}(R+\bar R,i(\bar R-R),1-|R|^2).
\label{invsp}
\ee
The approximate monopole fields in AdS are taken to be
\be
\Phi=iH{\bf n}\cdot \pauli, \quad\quad\quad
A_j=\frac{i}{2}(1-K)({\bf n}\times\partial_j{\bf n})\cdot \pauli,
\label{rat}
\ee
where $H(\rho)$ and $K(\rho)$ are real radial profile functions.
Regularity at the origin imposes the boundary conditions
$H(0)=0$ and $K(0)=1.$ 
The conditions at the spatial boundary of AdS follow from the
requirement that as $\rho\rightarrow 1$ then $|\Phi|\rightarrow 1$
and $D_j\Phi\rightarrow 0.$ This provides the boundary conditions
$H(1)=1$ and  $K(1)=0.$ 

A more general form for the gauge potential, using all the
symmetric tensors that can be computed from ${\bf n},$ appears to be
\be
A_j=\frac{i}{2}(1-K)({\bf n}\times\partial_j{\bf n})\cdot \pauli
+\frac{i}{2}P\,\partial_j{\bf n}\cdot \pauli
+\frac{i}{2}Q\,{\hat x}_j{\bf n}\cdot \pauli,
\ee
with additional profile functions $P(\rho)$ and $Q(\rho).$
However, $Q$ corresponds to an abelian gauge potential in the
reduced one-dimensional radial theory, and therefore may be set to
zero by a gauge transformation. Furthermore, there is a global
$U(1)$ symmetry that rotates the fields $K$ and $P,$ and this 
can be used to set $P$ to zero.

As $\Phi|_{\rho=1}=i{\bf n}\cdot \pauli,$  it is obvious
that the monopole charge $N,$ which is the winding number of the Higgs
field on the boundary two-sphere, is equal to the degree of the
rational map $R(z)$ that determines ${\bf n}$ via (\ref{invsp}).
It will be useful later to note the integral expression for the
degree
\be
N=\frac{1}{4\pi}\int
\bigg(
\frac{1+|z|^2}{1+|R|^2}\bigg|\frac{dR}{dz}\bigg|\bigg)^2
\, \frac{2i\,dzd\bar z}{(1+|z|^2)^2}
\label{n}
\ee
where the final factor in (\ref{n}) is simply the
standard area element on the two-sphere.
 
Substituting the approximate fields (\ref{rat}) into the energy (\ref{energy})
yields the expression
\be
E^{rat}=4\pi L\int_0^1\bigg\{
\frac{1}{(1-\rho^2)^2}(\rho^2 H'^2+2NH^2K^2)
+\frac{1}{16L^2}\bigg(2NK'^2+\frac{{\cal I}}{\rho^2}(K^2-1)^2\bigg)
\bigg\}(1+\rho^2)\,d\rho,
\label{ratenergy}
\ee
where ${\cal I}$ is defined as the following energy functional on
the space of rational maps
\be
{\cal I}=\frac{1}{4\pi}\int
\bigg(
\frac{1+|z|^2}{1+|R|^2}\bigg|\frac{dR}{dz}\bigg|\bigg)^4
\, \frac{2i\,dzd\bar z}{(1+|z|^2)^2}.
\label{bigi}
\ee

The functional (\ref{bigi}) is precisely the same quantity that
appears in the rational map approximation of Skyrmions \cite{HMS}, 
and in both situations this is the only contribution to the energy
that distinguishes between rational maps of the same degree $N.$
This proves that, within the rational map approximation, the minimal energy
monopole of charge $N$ in AdS has the same rotational symmetries
as the minimal energy Skyrmion with baryon number $N$ and massless
pions in Minkowski spacetime. 

Previous results on rational maps, obtained in the context of Skyrmions,
will therefore be of use later in this paper.
For example, 
an inequality that follows simply from (\ref{n}) and (\ref{bigi})
is ${\cal I}\ge N^2$ \cite{HMS}. 
 Furthermore,
for a large range of $N,$ 
including all $N\le 40,$ 
the rational maps that minimize 
${\cal I}$ have been computed numerically \cite{HMS,BS-full,BHS},
and their symmetries identified. Recall that a rational map is
symmetric under a group $G\in SO(3)$ if a spatial rotation
$g\in G,$ which acts on the Riemann sphere coordinate $z$ as
an $SU(2)$ M\"obius transformation, can be compensated by an
 $SU(2)$ M\"obius transformation acting on the target 
Riemann sphere coordinate $R.$ A rotation on the
target sphere corresponds to a gauge transformation, and hence a symmetry,
of the monopole. 

The monopole fields (\ref{rat}) are consistent with the 
static field equations only if the rational map is spherically symmetric.
The only spherically symmetric rational map is the 
(unique up to the action of $SU(2)$ M\"obius transformations) degree one map
$R=z.$ In this case ${\cal I}=N=1$ and the rational map approximation is
exact, with the expression (\ref{ratenergy}) for $E^{rat}$
reproducing the true monopole energy. In this case the ordinary differential
equations for the profile functions $H$ and $K,$ that follow from
the variation of (\ref{ratenergy}), agree with those appearing in the
previous investigations \cite{LS,LMS} of the single monopole in AdS.    

A numerical construction of the profile functions allows a 
computation of the 1-monopole energy as a function of the AdS
radius $L$. The results reveal that for $L\gtrsim \frac{1}{3}$
the energy is well-approximated by the formula
$E\approx 2\pi(1+\frac{2}{5}L^{-1}).$ In particular, the
energy is $E=2\pi\times 1.396$ for $L=1.$ In all the numerical
computations presented in this paper the value $L=1$ will be chosen
as a generic radius. The above result shows that for this radius 
the energy of the single monopole is increased by around $40\%$ from
the flat space limit, which should be sufficient to observe the 
phenomena that arise due to the curvature of AdS. In the analytic
approximations the dependence on $L$ will be retained, allowing the
qualitative behaviour with $L$ to be deduced.

\begin{figure}[ht]\begin{center}
\includegraphics[width=12cm]{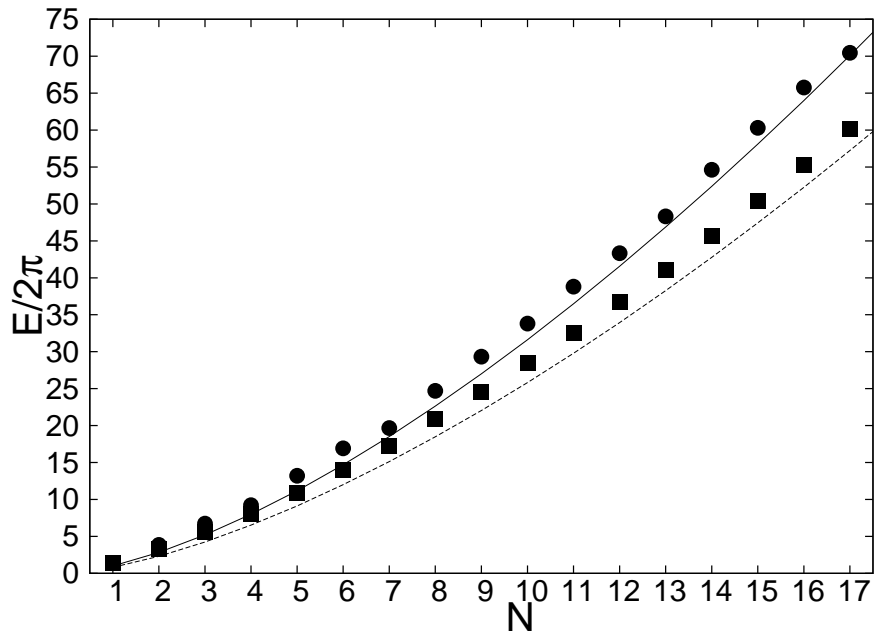}
\caption{The energy (in units of $2\pi$) for monopoles with
charge $N\le 17.$ Circles denote the values obtained from the 
rational map approximation and squares are the results of field
theory simulations. The solid curve is $N^{3/2}$ and the dashed curve
is the magnetic bag approximation.}
\label{fig-num}\end{center}\end{figure}

For each charge $N,$ using the minimal value of ${\cal I}$ in
(\ref{ratenergy}) and numerically computing the profile functions
$H$ and $K,$ produces an energy $E^{rat}$ which is an upper bound
on the true minimal energy of the $N$-monopole. 
The energies $E^{rat}$ (in units of $2\pi$) are plotted as the
circles in Figure~\ref{fig-num} for $1\le N\le 17.$

The rational map approximation may be related to the magnetic
bag approximation \cite{Bo,BT} by making a series of further
simplifications, as described below.
The magnetic bag approximation assumes that the fields take 
different forms inside and outside a bag of radius $\rho=\sigma.$
Inside the bag, that is $\rho\in[0,\sigma)$ the profile functions 
are taken to be $H=0$ and $K=1,$ so that the energy density
vanishes. For  $\rho>\sigma,$ which is the outside of the bag, 
the magnetic field is taken to be abelian by setting $K=0.$ 
With these simplifications the energy $E^{rat}$ becomes
\be
E^{rat}_{bag}=4\pi L\int_\sigma^1\bigg\{
\frac{\rho^2 H'^2}{(1-\rho^2)^2}
+\frac{{\cal I}}{16L^2\rho^2}
\bigg\}(1+\rho^2)\,d\rho,
\ee 
where the contribution from the surface of the bag has been ignored,
which means that $E^{rat}_{bag}$ is no longer guaranteed to 
be an upper bound for the true monopole energy.
The boundary condition on the surface of the bag is $H(\sigma)=0,$
and the energy minimizing profile function is easily found to be
\be
H=\frac{h(\sigma)-h(\rho)}{h(\sigma)-\pi},
\ee
where the function
\be 
h(\rho)=\frac{1}{\rho}-\rho+4\tan^{-1}\rho,
\ee
has been introduced.
The resulting energy depends on the bag radius $\sigma$ via
\be
E^{rat}_{bag}=4\pi\bigg\{
\frac{L}{h(\sigma)-\pi}+\frac{{\cal I}}{16L\sigma}(1-\sigma^2)\bigg\}.
\ee
The magnetic bag approximation is expected to be accurate
for large charge $N\gg L,$ when the bag
radius approaches the AdS boundary, that is, $\sigma=1-\epsilon$ with
$0<\epsilon\ll 1.$ To leading order in $\epsilon$
\be
E^{rat}_{bag}=2\pi\bigg(
\frac{3L}{\epsilon^3}+\frac{{\cal I}\epsilon}{4L}\bigg),
\ee
and the energy is minimized when
$
\epsilon=\sqrt{6L}/{\cal I}^{1/4},
$
to give the value 
\be
E^{rat}_{bag}=2\pi\sqrt{\frac{2}{3L}}{\cal I}^{3/4}
\approx 2\pi\sqrt{\frac{N^3}{L}}
\label{ratbag}
\ee
where the final approximation uses the fact that,
for a large range of $N,$ numerical results reveal \cite{BHS} that
${\cal I}\approx 1.3 N^2\approx (\frac{3}{2})^{2/3}N^2.$
In fact, the analysis in section \ref{sec-wall} suggests the
limiting behaviour ${\cal I}/N^2\rightarrow 1.21$ as $N\rightarrow \infty,$
which is consistent with the numerical results for large $N.$

The solid curve in Figure~\ref{fig-num} is the function $N^{3/2},$
obtained from the approximation (\ref{ratbag}) by setting $L=1.$
It can be seen that this approximation is a reasonable fit to
the numerical data (circles) and confirms the
superlinear growth of $E^{rat}$ with $N.$  
As discussed in the introduction, in AdS a superlinear growth of the
monopole energy with charge does not imply that the $N$-monopole 
is unstable to fragmentation into individual monopoles.

The original magnetic bag approximation \cite{Bo,BT} assumes that the angular
distribution is spherical. In terms of the rational map approximation
this corresponds to the assumption that the winding density
\be
\bigg(
\frac{1+|z|^2}{1+|R|^2}\bigg|\frac{dR}{dz}\bigg|\bigg)^2,
\label{nden}
\ee
appearing in the expression (\ref{n}) for the charge, is constant.
Of course, the only rational map for which this is true is the
degree one map $R=z,$ corresponding to the fact that there are
no spherical monopoles with $N>1.$ However, ignoring this fact
and assuming that the density (\ref{nden}) is constant, then it must
be equal to $N.$ In which case ${\cal I}=N^2,$ so the spherical
assumption is equivalent to approximating ${\cal I}$ by its lower bound
$N^2$.
With this additional simplification the energy $E^{rat}_{bag}$ in
(\ref{ratbag}) becomes the bag energy
\be
E_{bag}= 2\pi\sqrt{\frac{2N^3}{3L}}\,
\label{bag}
\ee
first obtained by Bolognesi and Tong \cite{BT}. 
The dashed curve in Figure~\ref{fig-num} is this magnetic bag energy.

The energy $E^{rat}$ of the rational map approximation is an upper
bound for the true monopole energy, and the bag energy $E_{bag}$ is
obtained by neglecting certain contributions to the energy and
assuming an idealized spherical distribution. It might therefore be
expected that the true monopole energy lies between these two 
approximations. In the following section numerical results 
will be presented that confirm this expectation, using
simulations of the full non-abelian nonlinear
field theory.

\section{Field theory simulations}\news
To perform numerical field theory simulations it is convenient 
to introduce Cartesian type coordinates ${\bf x},$ 
defined inside the unit ball, 
so that the sausage metric becomes
\be
ds^2=-\bigg(\frac{1+\rho^2}{1-\rho^2}\bigg)^2\,dt^2
+\frac{4L^2 \,d{\bf x}\cdot d{\bf x}}{(1-\rho^2)^2},
\ee
where $\rho^2={\bf x}\cdot {\bf x}\le 1.$

The numerical monopole solutions are obtained by performing a
simulated annealing energy minimization algorithm on the associated
static energy
\be
E=\int -\frac{1}{2}\mbox{Tr}\bigg\{
\frac{1}{8L}F_{ij}^2+\frac{L}{(1-\rho^2)^2}(D_i\Phi)^2\bigg\}
(1+\rho^2)\,d^3x.
\ee
The coordinates ${\bf x}$ are discretized on a regular lattice with
lattice spacing $dx=0.02,$ and the simulation grid contains all the
points of the lattice that satisfy ${\bf x}\cdot {\bf x}\le 1.$
Spatial derivatives are approximated using a second order finite difference
scheme and the energy is computed at points of the dual lattice.
On the boundary of the grid the vacuum expectation value of the
Higgs field is enforced $|\Phi|=1,$ and $\Phi$ is fixed at the 
sampled values of a continuum field with winding number $N.$
Explicitly, on the boundary $\Phi=i{\bf n}\cdot \pauli$ with 
${\bf n}$ given in terms of a degree $N$ rational map via (\ref{invsp}). 
In theory, all degree $N$ rational maps should provide equivalent
boundary conditions, as the winding number of the Higgs field is
the only gauge invariant quantity. However, on the lattice there
is likely to be a small bias against rational maps with extreme
angular derivatives.

\begin{table}[ht]
\centering
\begin{tabular}{|c|c|r|}
\hline
$N$ & $G$ &  $E/(2\pi)$ \\ \hline
1 & $O(3)$ & 1.39 \\ 
2 & $D_{\infty h}$ & 3.29 \\
3 & $T_d$ & 5.54 \\
4 & $O_h$ &  8.08 \\
5 & $D_{2d}$ &  10.94  \\
6 &  $D_{4d}$ &  14.01 \\
7 & $Y_h$ & 17.29   \\
8 &  $D_{6d}$ & 20.83   \\
9 &$D_{4d}$  & 24.55  \\
10 &$D_{4d}$  &  28.41 \\
11 & $D_{3h}$ & 32.49  \\
12 & $T_d$ & 36.71  \\
13 & $O$ & 41.08   \\
14 & $D_2$ & 45.65   \\
15 & $T$ & 50.33   \\
16 & $D_3$ & 55.29   \\
17 & $Y_h$ & 60.07  \\
\hline
\end{tabular}
\caption{The symmetry group and energy (in units of $2\pi$)
for monopoles with charge $N\le 17.$ }
 \label{tab-energies}
\end{table}

The final ingredient required for the numerical simulation is
an initial condition. The only requirement on the initial fields
is that the Higgs field on the boundary must take the form 
described above, namely $\Phi=i{\bf n}\cdot \pauli,$ with ${\bf n}$
determined by a particular rational map. The simplest possibility is
to take the initial fields to be
\be
\Phi=i\rho{\bf n}\cdot \pauli, \quad\quad
A_j=0.
\label{ic}
\ee
Note that in the continuum theory this initial field does not have finite
energy, because $D_j\Phi$ does not vanish on the boundary. 
However, this is not a problem on the lattice, and the gauge potential
evolves during the simulation, and in particular at the boundary, to
drive $D_j\Phi$ towards zero at the boundary.  
A more sophisticated initial condition could be used, for example
by starting with the fields of the rational map approximation, but
this is not necessary.

An indication of the numerical accuracy of the algorithm can
be obtained by computing the charge $N,$ of the final 
configuration, using the 
lattice version of (\ref{charge}). For the results
presented in this section 
the numerical charge is integer-valued to an accuracy of
within $0.1\%$ for $N\le 4,$ with the error rising to $1\%$ for
$N=17,$ which is the largest value considered.
 
The minimal energies obtained from field theory simulations
are presented in Table~\ref{tab-energies} and plotted as the 
squares in Figure~\ref{fig-num}. As expected, these energies lie between
the values predicted by the rational map and  
magnetic bag approximations. The rational map approximation is more
accurate for small values of $N,$ but the magnetic bag approximation
becomes increasingly accurate as $N$ increases.

\begin{figure}[ht]\begin{center}
\includegraphics[width=16cm]{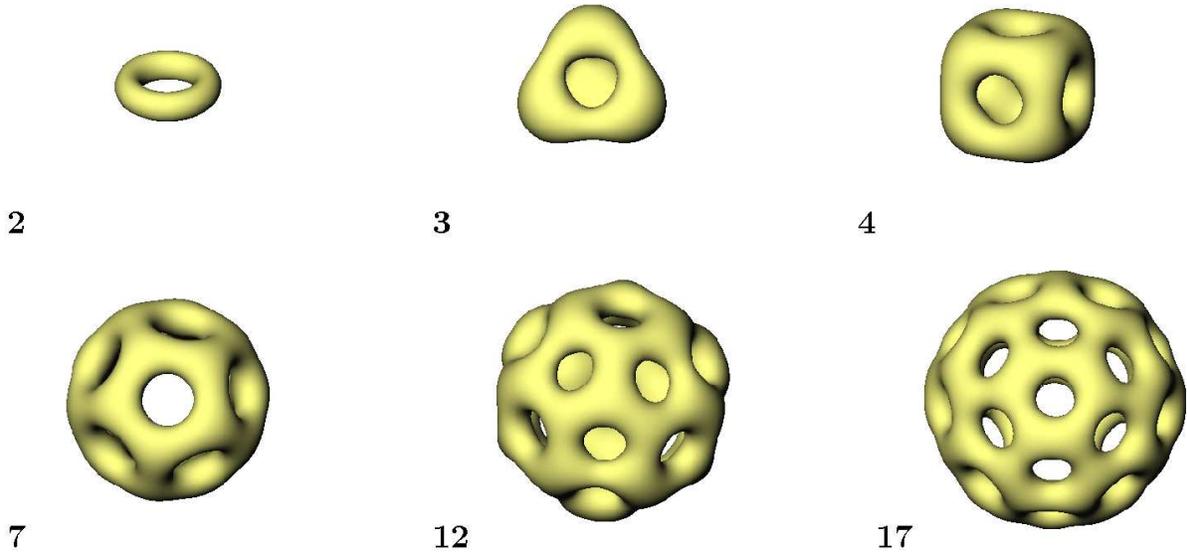}
\caption{Energy density isosurfaces for monopoles with
charges $N=2,3,4,7,12,17.$
Each of these monopoles has either a continuous or Platonic symmetry
and the plots are displayed to scale.
}
\label{fig-iso}\end{center}\end{figure}

To obtain these numerical results the rational map used in the 
initial (and boundary) condition is the ${\cal I}$ minimizing map for
each value of $N.$ The symmetry of this map, which is also found to be 
the symmetry of the final numerical solution, is listed in 
Table~\ref{tab-energies}.  
Energy density isosurfaces are displayed in Figure~\ref{fig-iso}
for a selection of monopoles with continuous or Platonic symmetries.
For $N>2$ the energy density is localized on the edges, and 
particularly the vertices, of a polyhedron. For the larger charges
shown in Figure~\ref{fig-iso} the pattern emerges of a polyhedron
with $2N-2$ faces, of which 12 are pentagons and the remainder are hexagons.
Such polyhedra are familiar from the study of fullerenes in carbon
chemistry, with the charge $17$ monopole providing the most famous example
of the truncated icosahedron, associated with the buckyball. 
The monopole energy density isosurfaces presented in Figure~\ref{fig-iso}
are qualitatively the same as those for Skyrmions, where the 
connection with fullerenes was first observed \cite{BS-full}.

Note that the difference between
the approximate rational map energy (the last expression in (\ref{ratbag}))
and the magnetic bag energy (\ref{bag}) is a factor $\sqrt{3/2}=1.22...$
This suggests that generically the energy of the rational map approximation
may be as much as $20\%$ above the true value. This is much larger than
the error in the rational map approximation to Skyrmions \cite{HMS}, which
is typically a few percent. This means that some caution must be exercised
when concluding that the symmetry predicted by the rational map approximation
is indeed the symmetry of the minimal energy monopole. However, the
fact that the numerical solutions, which share the same symmetries as the
rational maps, have energies that tend towards the magnetic bag values,
suggests that these solutions are strong candidates for minimal 
energy monopoles. 

Further evidence to support this claim is provided by considering
alternative rational maps. For example, if the axially symmetric
degree three rational map $R=z^3$ is used to provide the initial
(and boundary) conditions for the simulation then an axially symmetric
3-monopole is obtained with an energy $E=2\pi\times 5.64;$
this is a little larger than the energy $E=2\pi\times 5.54$ of the tetrahedrally
symmetric 3-monopole. Furthermore, symmetry breaking 
perturbations of axially symmetric maps have been simulated for $N=3$
and $N=4$ by using the rational map $R=z^N+\frac{1}{10}z^{N-1}$. 
In both cases the final numerical solution obtained is a Platonic monopole 
(tetrahedral for $N=3$ and cubic for $N=4$), making
it extremely likely that these are the minimal energy monopoles. 
It is difficult to perform
similar simulations using asymmetric maps for larger values of $N,$
both because of the dramatically increased simulation time required
in comparison to symmetric maps, and the fact that the perturbed map
must remain as the boundary map throughout the simulation, which
leads to increased angular derivatives as the charge increases.

Other symmetric maps that are not ${\cal I}$ minimizing have also
been used in simulations. For example, there is a degree five map
with octahedral symmetry \cite{HMS} that yields a numerical solution for
an octahedron with energy $E=2\pi\times 10.98$;
this is slightly larger than the energy $E=2\pi\times 10.94$ of the 
less symmetric 5-monopole with only $D_{2d}$ symmetry. 
There is also an icosahedrally symmetric degree eleven map \cite{HMS}
that produces an icosahedron with an energy $E=2\pi\times 32.85$, again
larger than the less symmetric  $D_{3h}$ 11-monopole with
energy $E=2\pi\times 32.49.$ It seems likely that these
additional solutions are saddle points.

The rational map approximation will predict the correct symmetry
of the minimal energy monopole providing the error in the
approximation is similar for competing local minima. One situation
in which this is clearly not the case is the flat space limit 
$L\rightarrow \infty$  of Minkowski spacetime, with its $4N$-dimensional 
BPS moduli space ${\cal M}_N.$ The rational map approximation (\ref{rat})
survives the limit to Minkowski spacetime, but it does not capture
the BPS moduli space. In this case, the rational map dependence of the
energy is not a physical property but rather an indication of the 
failure of the rational map approximation to accurately describe a
particular monopole in the moduli space. The rational map approximation
is at its most accurate for shell-like configurations, hence 
${\cal I}$ may be thought of as a measure of the deviation from
spherical symmetry of an $N$-monopole, which is unattainable for $N>1.$
For example, for $N=4$ the ${\cal I}$ minimizing map has cubic
symmetry and the fields (\ref{rat}) provide an approximation to
the exact cubic 4-monopole \cite{HMM}, which may be regarded as the
point in ${\cal M}_4$ which is the closest to a spherical 4-monopole.
In this example the energy of the rational map approximation turns
out to be $10\%$ above the BPS energy, and the error is obviously greater
than this in attempting to describe any other monopole in the 
moduli space ${\cal M}_4.$ Despite these limitations, the rational
map approximation may turn out to be useful in Minkowski spacetime,
as it provides explicit monopole fields that can be used in analytic 
approximations, or to provide initial conditions in any numerical 
computations involving monopoles. 
 
The magnetic bag approximation assumes that the Higgs field
vanishes throughout a ball. However, exact monopole solutions 
have a finite number of zeros of the Higgs field. In the rational
map approximation it is assumed that there is a single zero at the
origin, with multiplicity $N$ for an $N$-monopole. This is a
correct description of almost all the numerical monopole solutions
presented in this paper. The exceptions are the monopoles of
charges three, five and eleven, describing a tetrahedron, octahedron and
icosahedron respectively. For these monopoles, of which only
the $N=3$ example is minimal energy, the associated polyhedron has 
triangular faces and $N+1$ vertices. There are $N+1$ zeros of the
Higgs field associated with these vertices (but at a 
smaller distance from the origin) plus an additional zero of
the Higgs field at the origin with a negative multiplicity (termed
an anti-zero). This mirrors the situation in Minkowski spacetime,
where the same phenomenon has been discovered for the tetrahedral $N=3$
and octahedral $N=5$ examples \cite{HS3,Su4} using the Nahm transform. 
The $N=11$ icosahedral example has not been investigated in 
Minkowski spacetime because the Nahm data is not known in this case.

\begin{figure}[ht]\begin{center}
\includegraphics[width=10cm]{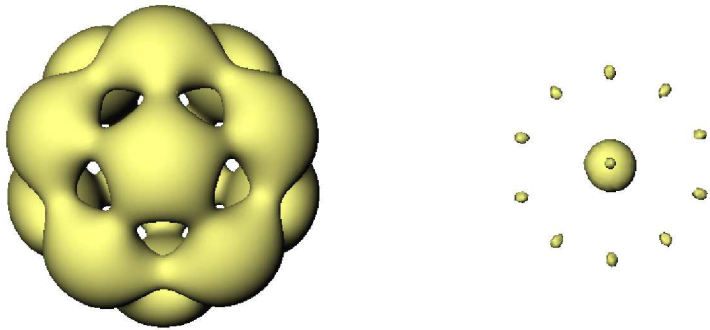}
\caption{
On the left is an energy density isosurface for the
icosahedral 11-monopole. On the right (to scale) is the associated
isosurface where $|\Phi|=0.06,$ indicating the positions of the
zeros of the Higgs field. There are 12 zeros on the vertices of an
icosahedron and an anti-zero at the origin.
}
\label{fig-icos}\end{center}\end{figure}

In Figure~\ref{fig-icos} the properties of the icosahedral
11-monopole in AdS are displayed. On the left is an energy density isosurface,
confirming that the energy density is maximal on the vertices 
of an icosahedron. On the right is a plot to highlight the positions of
the zeros of the Higgs field, by displaying 
an isosurface where $|\Phi|$ is small; in this particular
case $|\Phi|=0.06.$ This plot reveals that there are 
12 zeros of the Higgs field on the vertices of an icosahedron
 and an anti-zero at the origin. The surface around the anti-zero
appears much larger than that around the 12 zeros, 
which may indicate that the variation of the length of the Higgs 
field is reduced around the anti-zero. However, this conclusion is
not certain because of the visual distortion in size associated
with the AdS metric.

It has been suggested, at least in Minkowski spacetime, that there
may be different types of monopole bags depending upon whether 
the zeros of the Higgs field are located at the origin or on the shell
of the bag \cite{LW}. The numerical results in AdS show that there are
indeed two different types of monopole solutions characterized either
by $N$ zeros at the origin or $N+1$ zeros near the points of maximal
energy density, plus an anti-zero at the origin. In AdS it appears
that for $N>3$ the former type of monopole solution has  
lower energy.

\section{Monopole walls}\news\label{sec-wall}
In the large $N$ limit, monopoles in AdS can be studied by zooming to
the Poincar\'e patch of AdS. This results in the emergence of a monopole wall,
associated with the shell of the $N$-monopole on which the 
energy density is maximal.
Monopole walls were initially studied by Ward \cite{Wa} in Minkowski spacetime,
as novel domain walls in which the magnetic field 
along a line perpendicular to the wall 
tends to zero on one side of the wall and to a non-zero constant on 
the other side of the wall.
In contrast to standard domain walls, the fields are not independent
of the coordinates perpendicular to the wall, but are periodic with 
a spatial variation of the energy density and magnetic field. 

In Minkowski spacetime monopole walls have infinite energy per unit area.
However, in AdS the energy per unit area of a monopole wall is finite,
 and can be investigated using a magnetic bag style approximation \cite{BT}.
The magnetic bag approximation ignores the spatial distribution of the 
fields within the wall, and is therefore unable to make any 
predictions about the symmetry or lattice structure of the wall.
Numerical results in Minkowski spacetime \cite{Wa} are unable to shed any
light on this issue, as the symmetry is controlled by free parameters
in the solution, that are essentially inherited from the BPS moduli space
for finite charge.

A variant of the rational map approximation is introduced 
in this section, that is suitable for studying monopole walls in AdS.
This approximation is refined enough to address the spatial
distribution of the fields and suggests that the minimal energy per 
unit area is obtained from a monopole wall with a hexagonal lattice.
Numerical simulations are performed that support this conclusion, 
which is again in agreement with the Skyrme model, where a similar
hexagonal lattice exists \cite{BS-lat}.      

In most applications of holographic methods, the Poincar\'e patch
of AdS is the correct arena in which to apply the AdS/CFT correspondence.
The results of an analysis of monopole walls in AdS is therefore likely 
to be of some interest within the context of holography, and in 
particular for applications to condensed matter systems where 
magnetic fields play a prominent role.   

In the Poincar\'e patch the planar metric of AdS reads
\be
ds^2=\frac{r^2}{L^2}(-dt^2+dx^2+dy^2)+\frac{L^2}{r^2}dr^2,
\label{planar}
\ee
where $r\ge 0$ is the radial variable in the bulk, with 
three-dimensional Minkowski spacetime obtained at the 
ultra-violet boundary $r\rightarrow\infty.$

Working in a fundamental torus, ${\bb T}^2,$ in the $(x,y)$ plane,
the static energy in the volume ${{\bb{T}}^2\times {\bb{R}}}$
is given by
\be
E=\int -\frac{1}{4}\mbox{Tr}\bigg\{
\bigg(\frac{L}{r}\bigg)^4F_{xy}^2+F_{xr}^2+F_{yr}^2
+\bigg(\frac{r}{L}\bigg)^2(D_r\Phi)^2
+\bigg(\frac{L}{r}\bigg)^2\bigg((D_x\Phi)^2+(D_y\Phi)^2\bigg)
\bigg\}
\bigg(\frac{r}{L}\bigg)^2\,d^3x.
\label{fullwall}
\ee
The wall version of the rational map approximation has
the same form as (\ref{rat})
\be
\Phi=iH{\bf n}\cdot \pauli, \quad
A_j=\frac{i}{2}(1-K)({\bf n}\times\partial_j{\bf n})\cdot \pauli,
\label{wallrat}
\ee
where $H(r),K(r)$ are profile functions and
${\bf n}$ is again related to a Riemann sphere coordinate $R$
through (\ref{invsp}).
However, in this case $R(z)$ is a periodic function in ${\bb{T}}^2,$
where $z=x+iy$ is the complex coordinate in the plane. 

The boundary conditions on the profile functions are
again determined by regularity and finite energy to be
$H(0)=0,\ H(\infty)=1, \ K(0)=1, \ K(\infty)=0,$ where
the vacuum expectation value $|\Phi|\rightarrow 1$ as
$r\rightarrow\infty$ has been imposed.  

The ansatz (\ref{wallrat}) implies that $A_r=0$ and
\be
F_{xy}=2i(1-K^2)J{\bf n}\cdot \pauli,
\ee
where 
\be
J=\frac{1}{(1+|R|^2)^2}\bigg|\frac{dR}{dz}\bigg|^2.
\ee
The monopole charge in the fundamental torus is the degree of the map from the 
torus to the sphere and is given by
\be
N=\frac{1}{\pi}\int_{{\bb{T}}^2}J\,dx dy.
\ee
Using the approximate fields (\ref{wallrat}) and performing the
integration over ${\bb{T}^2}$ yields
\be
E=\int_0^{\infty}\bigg\{
A
\bigg(\frac{r}{L}\bigg)^2
\frac{1}{2}H'^2+N\pi\bigg(
K'^2+
\bigg(\frac{L}{r}\bigg)^2
4H^2K^2\bigg)
+\frac{2{\cal I}\pi^2}{A}
\bigg(\frac{L}{r}\bigg)^4
(1-K^2)^2\bigg\}
\bigg(\frac{r}{L}\bigg)^2\,dr,
\label{walle}
\ee
where $A$ is the area of the torus ${\bb{T}^2}$
and 
\be
{\cal I}=\frac{A}{\pi^2} \int_{{\bb{T}}^2}J^2\,dx dy,
\ee
is a quantity that is independent of the area $A.$

As $r\rightarrow\infty$ the abelian magnetic field
perpendicular to the wall is
\be
B=-\frac{1}{2}\mbox{Tr}(F_{xy}\Phi)=2J,
\ee
with magnetic flux per unit area
\be
B_\star=\frac{1}{A}\int_{{\bb{T}}^2}2J\,dx dy=\frac{2\pi N}{A}.
\label{bstar}
\ee
There is a one-parameter family of magnetic walls, labelled by
$B_\star,$ the magnetic flux per unit area at the ultra-violet boundary.
Using (\ref{walle}) and the definition (\ref{bstar}) the energy 
per unit area of the wall may be written as
\be
\frac{E}{A}=\int_0^{\infty}\bigg\{
\bigg(\frac{r}{L}\bigg)^2
\frac{1}{2}H'^2+\frac{B_\star}{2}\bigg(
K'^2+
\bigg(\frac{L}{r}\bigg)^2
4H^2K^2\bigg)
+\frac{B_\star^2{\cal I}}{2N^2}
\bigg(\frac{L}{r}\bigg)^4
(1-K^2)^2\bigg\}
\bigg(\frac{r}{L}\bigg)^2\,dr.
\label{wallpua}
\ee
Before presenting an analysis of the energy (\ref{wallpua}) it
is first worth discussing the wall analogue of the magnetic bag
approximation \cite{BT}. 

To the infra-red side of the wall
$r<\sigma,$ the profile functions 
are taken to be $H=0$ and $K=1,$ so that the energy density
vanishes. To the ultra-violet side of the wall $r>\sigma,$ 
the simplification is to assume that $K=0,$ so that the energy per unit 
area becomes
\be
\frac{E}{A}=\int_\sigma^{\infty}\bigg\{
\bigg(\frac{r}{L}\bigg)^2
\frac{1}{2}H'^2
+\frac{B_\star^2{\cal I}}{2N^2}
\bigg(\frac{L}{r}\bigg)^4\bigg\}
\bigg(\frac{r}{L}\bigg)^2\,dr,
\ee
where once again the contribution on the wall $r=\sigma$
has been ignored.

The energy minimizing profile function, satisfying the boundary
condition $H(\sigma)=0,$ is easily found to be
\be
H=1-\frac{\sigma^3}{r^3}.
\ee
The position of the wall is 
\be
\sigma=\sqrt{\frac{B_\star L^3\sqrt{{\cal I}}}{3N}}
\ee
with a resulting energy per unit area
\be
\frac{E}{A}
=\sqrt{\frac{4B_\star^3 L{\cal I}^{3/2}}{3N^3}}.
\ee

The final approximation required to reproduce the 
result of Bolognesi and Tong \cite{BT} is to assume
that the magnetic field is independent of the coordinates
parallel to the wall. 
This simplification implies that
$B=B_\star=2J$ is a constant, and hence the approximation
${\cal I}=N^2.$ This yields the wall position
and energy per unit area as
\be
\sigma=\sqrt{\frac{B_\star L^3}{3}},
\quad \quad
\frac{E}{A}
=\sqrt{\frac{4B_\star^3 L}{3}}.
\label{wallbag}
\ee

Returning to the energy (\ref{wallpua}), 
an explicit expression for $R(z),$ the map from the
torus to the sphere, is required to make further progress.
To obtain a wall with hexagonal symmetry
this map is chosen to be proportional to the Weierstrass elliptic
function $\wp(z)$ defined by the equation
\be
\wp'^2=4\wp^3-4.
\ee
This elliptic function has periods 
$\omega_1=\Gamma(\frac{1}{6})\Gamma(\frac{1}{3})/(2\sqrt{3\pi})$ 
and $\omega_2=\omega_1e^{i\pi/3},$ giving the 
required $60^\circ$ angle between the fundamental periods.
The precise form taken for the map is $R(z)=c\wp(z/a),$ where $a$ and $c$ are
real constants, with $a$ determined in terms of the area $A$ of the 
torus by $a^2=2A/(\sqrt{3}\omega_1^2).$
Recall that ${\cal I}/N^2$ is independent of $A$ (and hence $a$).

The elliptic function has a double pole in its fundamental parallelogram
and describes a map with degree $N=2.$  
Integrating over the torus
reveals that ${\cal I}/N^2$ is minimized for $c=0.70,$ when
it takes the value ${\cal I}/N^2= 1.21.$ Note that this value
is consistent with the large $N$ limit of the minimizing rational
maps discussed earlier.
Using the value ${\cal I}/N^2= 1.21$ allows the energy 
per unit area (\ref{wallpua}) of the elliptic map approximation 
to be calculated by computing the
minimizing profile function. 

There is a scaling isometry of the metric (\ref{planar}) that
relates monopole wall solutions with different values of the
magnetic flux per unit area. Explicitly, the metric is 
invariant under the scaling 
$x\mapsto \lambda^{-1}x, \
y\mapsto \lambda^{-1}y, \
t\mapsto \lambda^{-1}t, \
r\mapsto \lambda r
$
which yields $B_\star\mapsto \lambda^2 B_\star$ and 
$E/A\mapsto \lambda^3 E/A.$ This scaling symmetry may be used
to restrict the computations to a convenient positive value
of $B_\star.$

Numerical field theory simulations allow a computation of
the monopole wall energy (\ref{fullwall}) using a simulated
annealing algorithm similar to that discussed in the previous
section. It is convenient to perform the simulations over
two copies of the torus, so that a rectangular domain
$(x,y)\in[0,a\omega_1]\times[0,\sqrt{3}a\omega_1]$ may be
used with periodic boundary conditions. The numerical grid
contained $52\times 90$ grid points to cover each rectangle.

It is numerically more efficient if all the spatial coordinates take
values in a finite range, hence the simulations use the 
variable $u=r/(1+r),$ taking values in the unit interval. 
This interval is covered with $50$ grid points, so the total
grid contains $52\times 90\times 50$ points. The initial conditions
are taken from the elliptic map approximation. 

\begin{figure}[ht]\begin{center}
\includegraphics[width=16cm]{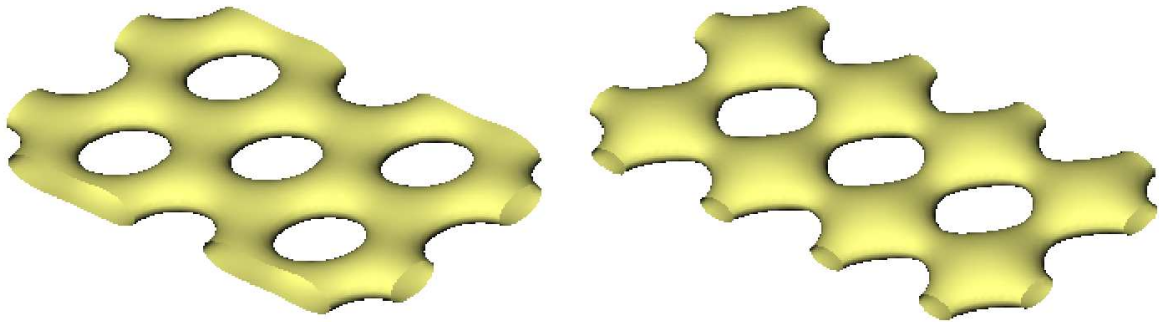}
\caption{Energy density isosurfaces for
monopole walls with $B_\star=10.$ On the left is the 
hexagonal wall and on the right is the square wall, which
has a slightly higher energy.}
\label{fig-lattice}\end{center}\end{figure}

In Figure~\ref{fig-lattice} the image on the left 
is an energy density isosurface for the numerical solution with $B_\star=10,$ 
where the hexagonal structure of the lattice is clearly visible.
As the region
contains two copies of the torus then the charge in the displayed
region is $N=4.$ 

The energy per unit area of this hexagonal wall 
is computed to be $E/A=39.1.$ 
For comparison, the energy per unit area of the
elliptic map approximation is $E/A=42.7$ and
for the magnetic bag style approximation is $E/A=36.5.$
As in the case of monopoles in global AdS, this demonstrates that 
the magnetic bag style approximation provides a lower bound on
the energy, to complement the upper bound of the elliptic map approximation. 

\begin{figure}[ht]\begin{center}
\includegraphics[width=10cm]{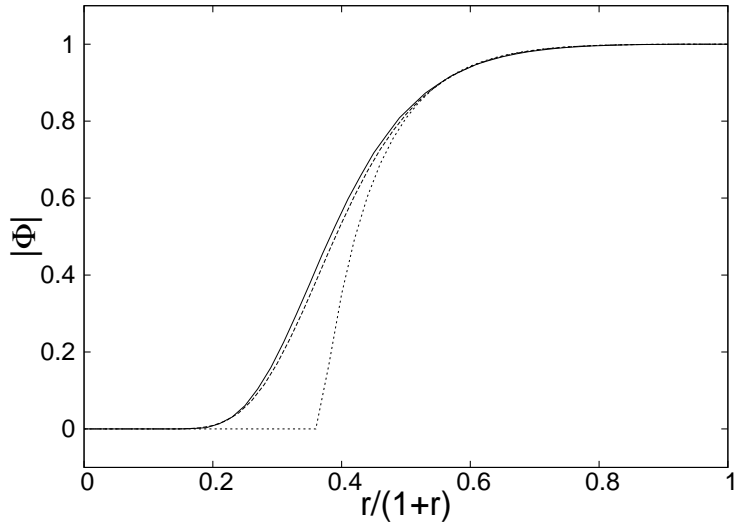}
\caption{$|\Phi|$ as a function of $r/(1+r).$
The dashed curve is the elliptic map approximation, the
dotted curve is the magnetic bag style approximation, and the solid
curve is the result from field theory simulations.
}
\label{fig-wallpro}\end{center}\end{figure}

In Figure~\ref{fig-wallpro} the length of the Higgs field 
$|\Phi|$ is plotted as a function of $r/(1+r)$
for the case $B_\star=1.$ 
The dashed curve is the elliptic map approximation and the 
dotted curve is the magnetic bag style approximation. The solid
curve is the result from field theory simulations, plotted along
a generic line perpendicular to the wall. It can be seen that
the elliptic map approximation is in excellent agreement with the
field theory simulations, and the magnetic bag approximation
provides a good description everywhere except around the actual wall
itself.

A monopole wall with a square lattice can be obtained by using 
the Weierstrass elliptic
function $\widetilde\wp(z)$ defined by the equation
\be
\widetilde\wp'^2=4\widetilde\wp^3-4\widetilde\wp,
\ee
which has periods 
$\widetilde\omega_1=\Gamma(\frac{1}{4})^2/(2\sqrt{2\pi})$ and 
$\widetilde\omega_2=i\widetilde\omega_1,$ producing a $90^\circ$ angle between
the fundamental periods.  

Taking a map of the form $R(z)=\widetilde c\widetilde\wp(z/\widetilde a),$
with $\widetilde a^2=A/\widetilde \omega_1^2,$
reveals that ${\cal I}/N^2$ is minimized for $\widetilde c=1.00,$ when
it takes the value ${\cal I}/N^2= 1.30.$ 
This is greater than that of the hexagonal lattice, and
supports the view that a hexagonal architecture produces
minimal energy. 

Further evidence is provided by using the approximate fields 
of the square wall
as initial conditions in the field theory simulations.
To facilitate a comparison with the simulations of the
hexagonal wall, two copies of the torus are again taken, 
by using the rectangular domain
$(x,y)\in[0,\widetilde a\widetilde \omega_1]\times
[0,2\widetilde a\widetilde \omega_1]$, covered by 
$50\times 100$ grid points.  
In Figure~\ref{fig-lattice} the image on the right 
is an energy density isosurface for the resulting 
numerical solution with $B_\star=10.$ 
The energy per unit area of this square wall is
$E/A=39.5,$ which is slightly greater than the
value $E/A=39.1$ for the corresponding hexagonal wall.

The connections between monopoles and Skyrmions suggests a
qualitative understanding of the preference for a hexagonal
monopole wall, since it is known that this is the minimal 
energy form for a wall in the Skyrme model \cite{BS-lat}. 
It seems that the same generic structure arises when both types of
soliton overlap, even though the source
of the soliton attraction is different in the two theories;
there being an attractive force between two solitons in the
Skyrme model and an attractive force between a soliton and
the origin of AdS in the case of monopoles.

\section{Conclusion}
Motivated by applications in holographic theories, a detailed
analytic and numerical study has been performed for monopoles
in AdS, and the associated monopole walls that arise in the large charge
limit. An approximation has been introduced, using rational maps
between Riemann spheres, and it has been confirmed that this provides
a reasonable description of the fields when compared with 
the results from field theory simulations. 
The results of these simulations also confirm
that the magnetic bag approximation yields energies that are close
to the true monopole energies, even for reasonably small values of
the charge. In all the numerical computations presented in this paper
the value $L=1$ has been used as a generic choice for the AdS radius.
It might be interesting to extend the computations to other values
of $L,$ or equivalently to a range of vacuum expectation values of the 
Higgs field. 

The arguments presented in \cite{BT} suggest that, 
for certain regions of parameter space,
the monopole wall (with its lattice structure) 
may be favoured over the more conventional magnetically charged black
hole, which is homogeneous at the ultra-violet boundary. 
It would be interesting
to investigate this issue by extending the methods described in this paper,
for a fixed AdS background, to the situation including gravitational
backreaction and finite temperature.

Finally, the connection between monopoles and Skyrmions 
in Minkowski spacetime has 
been enhanced by considering monopoles in AdS. 
A natural extension would therefore be to study Skyrmions
in AdS, to make a comparison with the results presented
here for monopoles. The rational map approximation 
easily extends to Skyrmions in AdS and yields the same 
functional on the space of rational maps. This would seem
to imply that monopoles and Skyrmions in AdS have a
similar form. However, there is a caveat to this conclusion.
It appears that, as far as Skyrmions are concerned, 
the curvature of hyperbolic space plays
a similar role to that of a pion mass \cite{AS}.
It is known that Skyrmions with massive pions, in Minkowski spacetime,
 are shell-like (and hence described by the rational
map approximation) only for sufficiently low baryon numbers, and
take a different form, including clusters, above a critical value 
\cite{BS-pm,BMS}. It may therefore be possible that Skyrmions
in AdS could take more exotic forms than monopoles, such as
clusters or multiple shells. 
It might be interesting to explore these possibilities.

\section*{Acknowledgements}
Many thanks to Gary Gibbons, Ruth Gregory, Derek Harland, 
Mukund Rangamani, Simon Ross, David Tong, Claude Warnick and
Marija Zamaklar for useful discussions.
I also thank the anonymous referee for helpful comments and
acknowledge STFC and EPSRC for grant support.

\end{document}